\documentclass[prl,showpacs,preprintnumbers,amsmath,amssymb,tightenlines,twocolumn]{revtex4}

\usepackage{graphicx}
\usepackage{dcolumn}
\usepackage{bm}
\usepackage{epsfig}

\newcommand{\ssection}[1]{{\noi  \it #1:}}
\newcommand{\bra}[1]{\langle\,{#1}\, |}
\newcommand{\ket}[1]{|\,{#1}\,\rangle}
\newcommand{\braket}[2]{\mbox{$\langle\,{#1}\, | \,{#2}\,\rangle$}}

\newcommand{\sub}[2]{{#1}_{\mbox{\!\! \scriptsize #2}}}

\newcommand{\bv}[1]{\mathbf{ #1 }}

\def\noi{\noindent}
\def\beq{\begin{equation}}
\def\eeq{\end{equation}}

\def\CR{\nonumber\\[0.15cm]}

\def\figurewidth{0.99}
\newcommand{\rref}[1]{ref.~\cite{#1}}
\newcommand{\fref}[1]{Fig.~\ref{#1}}
\newcommand{\frefp}[2]{Fig.~\ref{#1}~(#2)}

\newcommand{\eref}[1]{Eq.~(\ref{#1})}

\newcommand{\cref}[1]{chapter~\ref{#1}}

\newcommand{\Cref}[1]{Chapter~\ref{#1}}

\newcommand{\bref}[1]{(\ref{#1})}

\usepackage{ulem}  
\usepackage{color}

\begin{document}

\title{Quantum Zeno suppression of dipole-dipole forces}
\author{S.~W\"uster}
\affiliation{Max Planck Institute for the Physics of Complex Systems, N\"othnitzer Strasse 38, 01187 Dresden, Germany}
\affiliation{Department of Physics, Bilkent University, 06800 {\c C}ankaya, Ankara, Turkey}
\affiliation{Physics Department, Indian Institute of Science Education and Research, Bhopal, Madhya Pradesh 462 023, India}
\email{sebastian@iiserb.ac.in}
\begin{abstract}
We consider inter-atomic forces due to resonant dipole-dipole interactions within a dimer of highly excited Rydberg atoms, embedded in an ultra-cold gas.
These forces rely on a coherent superposition of two-atom electronic states, which is destroyed by continuous monitoring of the dimer state through a detection scheme utilizing controllable interactions with the background gas atoms. We show that this intrinsic decoherence of the molecular energy surface can gradually deteriorate a repulsive dimer state, causing a mixing of attractive and repulsive character.
For sufficiently strong decoherence, a Zeno-like effect causes a complete arrest of interatomic forces. We finally show how short decohering pulses can controllably  redistribute population between the different molecular energy surfaces.
\end{abstract}

\maketitle

\ssection{Introduction}
%
Through the extreme properties of Rydberg states \cite{book:gallagher}, ultracold Rydberg physics allows the study of chemical phenomena in hitherto unavailable realms.
Examples are homo-nuclear molecules bound over enormous distances and possessing a permanent dipole moment \cite{greene:ulr_mols,Bendkowsky:rydmol,Liu:borromeanrydbergtrimers,overstreet:csdimers,Butscher:amolcoh,Bendkowsky:internalquantrefPRL,weibin:permdipmopmscience,gaj:molspecdensshift,Gaj:Hybridization}, control of chemical reactions \cite{wang:control_chemreact} and non-adiabatic "nuclear" dynamics involving conical intersections \cite{yarkony2001conical,domke:yarkony:koeppel:CIs} over vastly inflated length scales \cite{wuester:CI,leonhardt:switch,leonhardt:unconstrained}. In all the latter cases, the role of chemical nuclei is taken by entire atoms including their valence electron and the motion of these is governed by Born-Oppenheimer (BO) surfaces as in molecular physics, that however arise in a simpler manner through just resonant electronic dipole-dipole interactions. Most importantly interactions have long ranges $\sim{\cal O}[10{\mu}$m$]$.

Since neighboring "nuclei" in Rydberg Chemistry are thus many orders of magnitude farther apart than in usual molecules, we enter a previously inaccessible regime where 
these can be coupled to \emph{separate}, strongly coupled perturbing environments. This enables novel physics through the loss of coherence between electron configuration basis states that make up a BO surface, thus \emph{instrinsic} to the energy surface. Previous studies of decoherence in molecules dealt instead with the loss of coherence between vibrational states on a given surface \cite{Schlesinger:molvibdecoh,lienau:vibcoh:jpc,Wallentowitz:vibcoh:JPB} or between different BO surfaces \cite{jasper:truhlar:elecdecoh}. In these cases decoherence is  due to coupling to the environment represented by additional molecular degrees of freedom of the molecule or a solvent.

A natural candidate for a quite different perturbing environment in the case of Rydberg chemistry is the cold background gas in which Rydberg molecules or aggregates are typically created. Exploiting the gas will additionally allow dynamical control of decoherence channels \cite{schoenleber:immag}, in contrast to the typical situation encountered in molecular physics. 

\begin{figure}[htb]
\centering
\epsfig{file=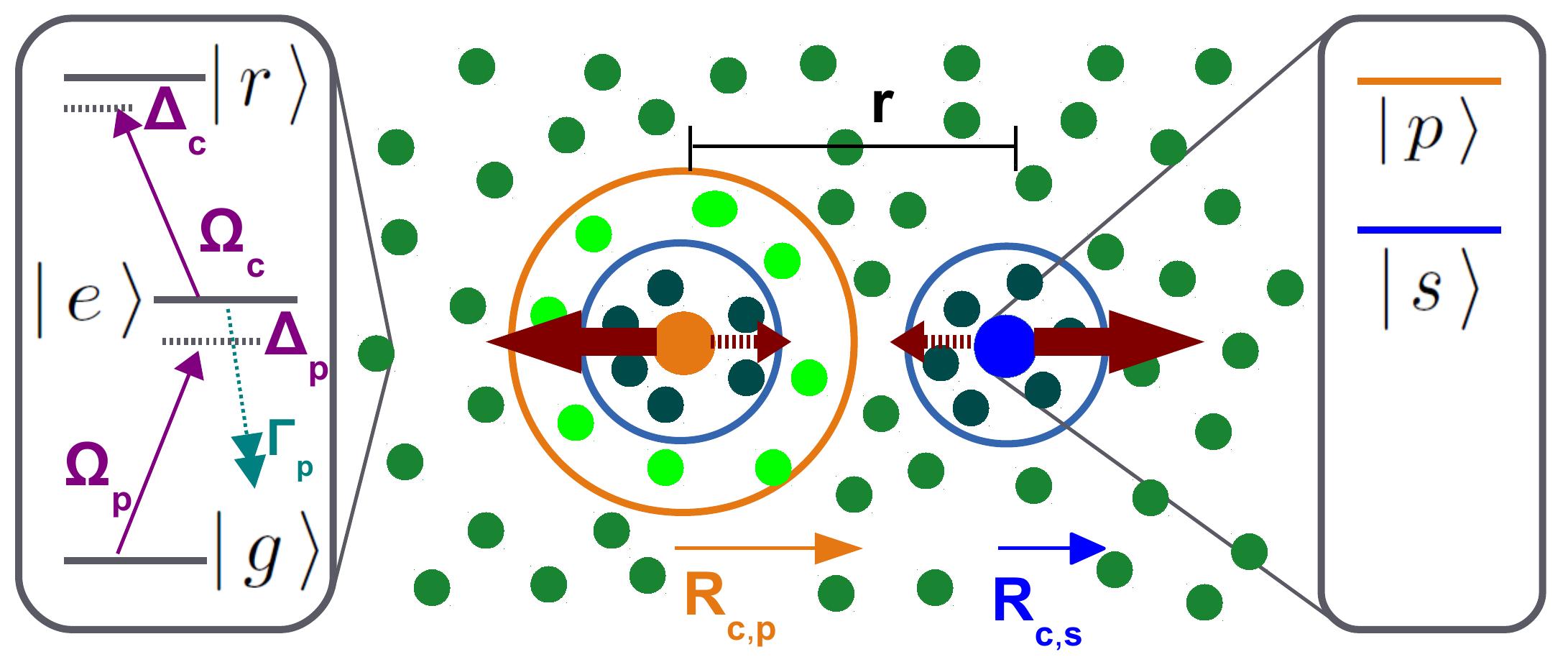,width= \figurewidth\columnwidth} 
\caption{(color online) Sketch of embedded Rydberg dimer. The dimer atoms (blue/orange) initially experience repulsive forces (solid brown arrows) due to resonant dipole-dipole interactions. Due to decohering interactions  with 
the background atomic cloud (green), these interactions gradually become partially attractive (dashed brown arrows). For annotations on level diagrams and critical radii $R_{c,s/p}$ see text.
\label{embedding_setup}}
\end{figure}

Here we thus consider a dimer of two dipole-dipole interacting Rydberg atoms, immersed in the background gas of ground-state atoms. Due to the exaggerated interactions of atoms in Rydberg states \cite{book:gallagher}, the dimer constituents are strongly coupled even when their separation $r$ vastly exceeds the mean ground-state atom spacing $d$, as sketched in \fref{embedding_setup}. Rydberg excitations within a background gas have been realized, keeping many of the properties of Rydberg states preserved despite the immersion \cite{heidemann:rydexcBEC,viteau:rydexclattice,balewski:elecBEC,gaj:molspecdensshift,karpiuk:imagingelectrons,niederpruem:molion,thaicharoen:trajectory_imaging,thaicharoen:dipolar_imaging,celistrino_teixeira:microwavespec_motion,Faoro:VdWexplosion_PRA}. Importantly the background offers means to probe the embedded Rydberg system \cite{olmos:amplification,guenter:EIT,guenter:EITexpt,mukherjee:phaseimp,karpiuk:imagingelectrons} and to  engineer controllable decoherence \cite{schoenleber:immag,schempp:spintransport,schoenleber:thermal}. 

In our system, the molecular states governing the dimer are simple coherent superpositions $\ket{\sub{\varphi}{rep/att}}=(\ket{ps} \pm (\ket{sp} )/\sqrt{2}$ of two electronic Rydberg states $\ket{s}$ and $\ket{p}$ of the constituent atoms. The subscripts indicate the repulsive or attractive character of these simple BO surfaces.
The background gas can now be used to infer the location (motion) and electronic state of these two Rydberg atoms as discussed in \rref{schoenleber:immag}. This allows to distinguish the two constituents of the superposition $\ket{ps}$ and $\ket{sp}$.
We show that the resulting measurement induced decoherence disrupts the superpositions on which the BO surfaces rely, changing the character of internuclear forces from repulsive to partially attractive. Earlier studies on decoherence of dipole-dipole interaction in Rydberg gases did not consider the effect on atomic motion, nor could a dipole-dipole interacting system and a decohering environment be distinguished \cite{kutteruf:coherence_dipdip_exp,Anderson:dephasing_energy_transf,Zhou:evolution:dipdip,schoenleber:thermal}.

We further show that dipole-dipole acceleration in the dimer can be brought to a complete arrest, furnishing a quantum Zeno effect \cite{misra:quantumzeno,Kofman:antizeno} for motional dynamics with experimentally accessible parameters. Finally, we demonstrate how short pulses of strong environment coupling (decoherence) can be exploited to shuffle population between BO-surfaces, in a further application of quantum state engineering through decoherence \cite{Tiersch:Decoherence:magnetoreception,schempp:spintransport,cenap:dissipbind,Lemeshko:dissipbind}.
 
\ssection{Scheme and model}
%
Consider a Rydberg dimer with inter-atomic separation $r$ embedded in a cold atom cloud of $M$ background atoms, as sketched in \fref{embedding_setup}. 
For the compound dimer, we allow only two electronic pair states $\ket{\pi_1}\equiv\ket{ps}$, with the first atom in $\ket{p}=\ket{\nu p}$ and the second in $\ket{s}=\ket{\nu s}$, and the reverse 
$\ket{\pi_2}\equiv\ket{sp}$. Here $\nu$ is the principal quantum number and angular momentum $l=0,1$ are denoted by $s$, $p$. 
The background atoms are initially prepared in the electronic ground state $\ket{g}$, and their positions could be random or arranged in a regular fashion. These atoms are coupled to two laser fields. One, with Rabi frequency $\Omega_p$ and detuning $\Delta_p$ drives transitions from $\ket{g}$ to a short-lived intermediate state $\ket{e}$. A second couples further from there to a third level, a Rydberg state $\ket{r}=\ket{\nu' s}$, 
with Rabi frequency $\Omega_c$ and detuning $\Delta_c$. The state $\ket{e}$ spontaneously decays with rate $\Gamma_p$ to $\ket{g}$. Overall we realize electromagnetically induced transparency (EIT)~ in the ladder configuration \cite{fleischhauer:review,friedler:longrangepulses,Mohapatra:coherent_ryd_det,mauger:strontspec,Mohapatra:giantelectroopt,schempp:cpt,sevincli:quantuminterf,
parigi:interactionnonlin}. 

We have shown in \rref{schoenleber:immag} based on \cite{guenter:EIT,guenter:EITexpt} how interactions between the dimer atoms and background atoms in state $\ket{r}$ allow one to infer which dimer atom is in the $\ket{p}$ state, thereby also providing controllable decoherence in the electronic state space spanned by $\ket{\pi_1}$, $\ket{\pi_2}$. While \rref{schoenleber:immag} was based on parameters where all atomic motion could be ignored in a frozen gas regime, we now focus explicitly on the effect of this decoherence on the dipole-dipole induced \emph{relative motion} of the Rydberg dimer. The Hamiltonian for only the dimer is 
\begin{align}
\hat{H}&= -\frac{\hbar^2 \nabla_{r}^2}{m} + \hat{H}_{dd}, \:\: \hat{H}_{dd} = W_{12}(r) \ket{\pi_1}\bra{\pi_2} + \mbox{c.c} ,
\label{Hdim}
\end{align}
describing two atoms of mass $m$ with relative co-ordinate $r$ and dipole-dipole interactions $W_{12}(r)=\mu^2/r^3$ ($W_{ii}\equiv0$). Anticipating the de-cohering effect of the background atoms described in \cite{schoenleber:immag} onto these dipole-dipole interactions, we model our system in terms of a density matrix $\hat{\rho}=\sum_{n,m}\int dr dr' \rho(r,r')_{nm} \ket{r}\otimes \ket{\pi_n}  \bra{r'}\otimes \bra{\pi_m}$, the elements of which follow the van-Neumann evolution equation
\begin{align}
&\sub{\dot{\rho}(r,r')}{nm}= -\frac{i}{\hbar}\big[-\frac{\hbar^2 }{2m}(\nabla_{r}^2 - \nabla_{r'}^2) \sub{\rho(r,r')}{nm} 
\CR
&+  \sum_{k}  (W_{nk}(r) \sub{\rho(r,r')}{km}  - W_{km}(r') \sub{\rho(r,r')}{nk} ) \big]
\CR
&+\left(i \frac{{\Delta}E}{\hbar}-\frac{\gamma}{2} \right) (1-\delta_{nm})\sub{\rho(r,r')}{nm}
.
\label{motionalmasterequation}
\end{align}
A derivation of \eref{motionalmasterequation} is given in the supplemental material \cite{sup:info}, where we also formally define the "position eigenstates" $\ket{r}$ used in the definition of $\hat{\rho}$.
The last row of \eref{motionalmasterequation} contains disorder (an energy shift) ${\Delta}E$ and dephasing $\gamma$ in the dimer electronic state space. Both are tuneable through the atomic and optical  
parameters $\Omega_{p/c}$,  $\Delta_{p/c}$, $\Gamma_p$ and interactions of dimer atoms in $\ket{s/p}$ with background gas atoms in $\ket{r}$ as described in \cite{schoenleber:immag,sup:info}.

The dephasing $\gamma$ arises because state dependent light absorption by the background gas allows one to experimentally distinguish the aggregate states $\ket{\pi_{1,2}}$. Briefly, interactions cause a breakdown of EIT within a critical radius $\sub{R}{c,s/p}$ around an $s/p$ impurity as shown in \fref{embedding_setup}, causing absorption shadows of the corresponding size. When $\sub{R}{c,s}\neq\sub{R}{c,p}$, observation of the shadow sizes allows discrimination of $\ket{\pi_{1,2}}$, for further details we refer to \cite{schoenleber:immag,sup:info}. The shadow centres correspond to the locations of the dimer atoms, so the same mechanism allows the observation of dimer separation $r$.

Since the Rabi frequencies $\Omega_{p/c}(t)$ can vary in time, also $\gamma$ can vary in time, hence we will refer to it as \emph{controllable decoherence}. Disorder ${\Delta}E$ originates from possibly different local environments of background atoms around each aggregate atom. 

In the following we explicitly consider atomic states $\ket{s}=\ket{43s}$, $\ket{p}=\ket{43p}$ and $\ket{r}=\ket{38s}$ for $^{87}$Rb. The resulting inter-atomic interactions can be found in \cite{schoenleber:immag}. For the parameters used, the disorder ${\Delta}E$ will be negligible and is hence set to zero.

\begin{figure}[htb]
\centering\epsfig{file=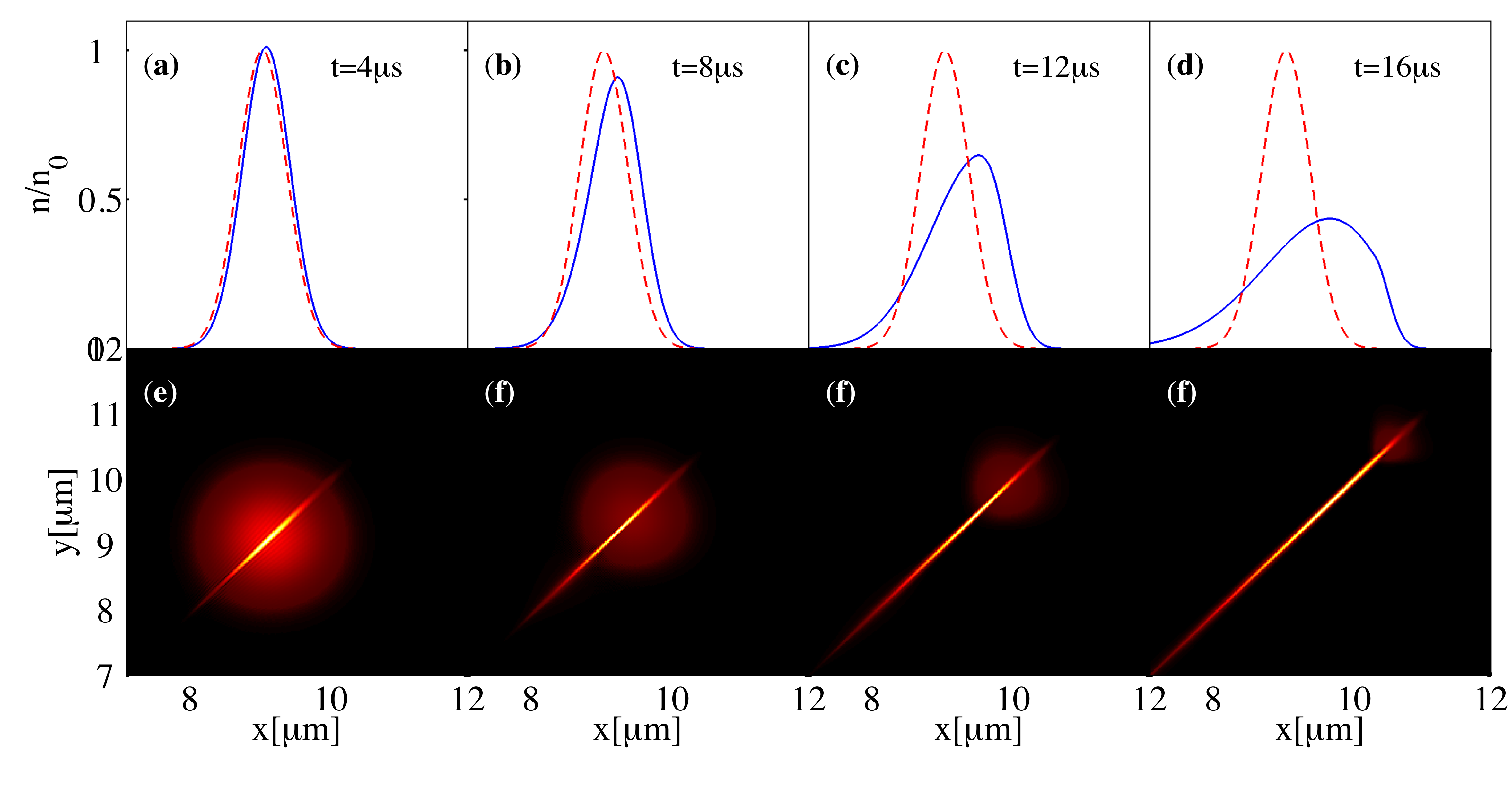,width= \figurewidth\columnwidth} 
\caption{(color online) Continuous decoherence turns an initially repulsive dipole-dipole interaction gradually into a mixture of attractive and repulsive dynamics, according to \eref{motionalmasterequation} for $\gamma=0.2$ MHz, $r_0=9$ $\mu$m,  $\sigma=0.5$ $\mu$m. (a-d) Probability density $n(r)=\sum_k  \sub{\rho(r,r)}{kk}$ of the relative coordinate at the indicated time (blue) and initial time, $t=0$ (red-dashed). (e-h) Corresponding density matrix $\sub{\rho(x,y)}{11}$ at the same times as the upper panels.
\label{results_gam}}
\end{figure}

\ssection{Decohering dipole-dipole interactions}
%
The dipole-dipole interaction Hamiltonian $\hat{H}_{dd}$ has two eigenstates $\ket{\sub{\varphi}{rep/att}}=(\ket{\pi_1} \pm (\ket{\pi_2} )/\sqrt{2}$, with repulsive and attractive potentials $\sub{U(r)}{rep/att}=\pm\mu^2/r^3$ \cite{cenap:motion,wuester:cradle,moebius:cradle}. 
We now investigate the effect of decoherence $\gamma$ on a dimer initialised in the repulsive state 
\begin{align}
\hat{\rho}(t=0) =\ket{\phi_0}\ket{\sub{\varphi}{rep}}\bra{\sub{\varphi}{rep}}\bra{\phi_0}
\label{rhoini}
\end{align}
through numerical solutions of \bref{motionalmasterequation} \cite{xmds:paper,xmds:docu}, as shown in \fref{results_gam}. In the expression above $\braket{r}{\phi_0}=\phi_0(r)={\cal N}\exp{[-(r-r_0)^2/(2\sigma^2)]}$ represents the initial wave function for the relative co-ordinate $r$, normalized to $1=\int dr |\phi_0(r)|^2$ via ${\cal N}$. 

For short times $t\lesssim8{\mu}s$ we see repulsive acceleration as expected. However the dephasing terms in \eref{motionalmasterequation} are gradually destroying the quantum coherence between $\ket{\pi_{1}}$ and $\ket{\pi_{2}}$, evolving the dimer state towards the incoherent mixed state $\hat{\rho}_M=[\ket{\pi_1}\bra{\pi_1} + \ket{\pi_2}\bra{\pi_2} ]/2$. The latter can also be written as $\hat{\rho}=[\ket{\sub{\varphi}{rep}}\bra{\sub{\varphi}{rep}} + \ket{\sub{\varphi}{att}}\bra{\sub{\varphi}{att}} ]/2$, thus this process will gradually populate the attractive potential. This can be seen in \fref{results_gam} at later times, where the dimer now is attractive with some probability. We also see that while the repulsively moving part of the system has largely preserved the initial spatial phase coherence ($\rho(r,r')>0$ for $|r-r'| \lesssim \sigma$), the attractively evolving part of the system has lost its phase coherence with the remainder, as it was created through a incoherent process.

\begin{figure}[thb]
\centering\epsfig{file=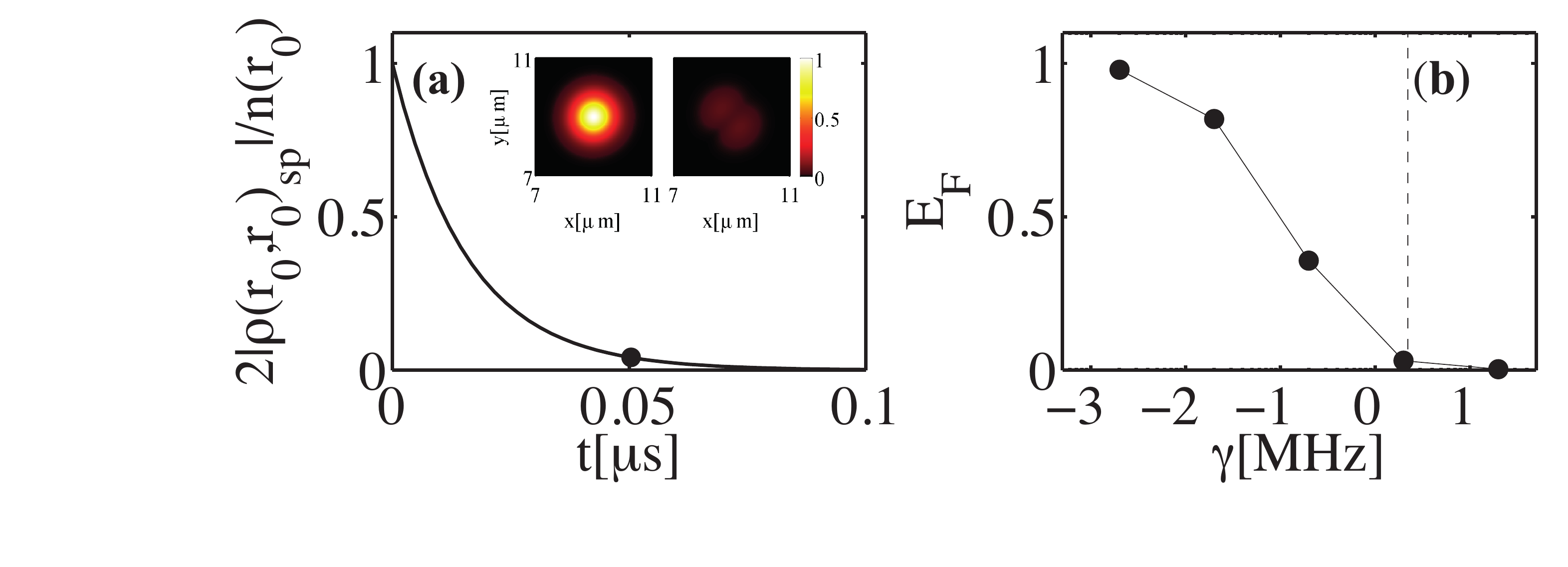,width= \figurewidth\columnwidth} 
\caption{(color online) (a) Rapid loss of coherence between $\ket{\pi_{1}}$ and $\ket{\pi_{2}}$ after initiating controllable decoherence with $\gamma=10$. The left [right] inset shows the spatial coherence $\sub{\rho(x,y)}{11}$ [$\sub{\rho(x,y)}{12}$] at the time indicated by ($\bullet$).
(b) Final kinetic energy of the dimer $E_f$ as a function of decoherence rate $\gamma$. We show the normalized quantity $\bar{E}_f = E_f/E_0$, where
$E_0$ is the final kinetic energy for $\gamma=0$. The dashed line indicates $W_{12}(r_0)$. Parameters other than $\gamma$ are as in \fref{results_gam}.
\label{zeno_arrest}}
\end{figure}

\ssection{Zeno arrest of motion}
%
For de-coherence rates $\gamma$ small compared to the initial dipole-dipole interaction $W_{12}(r_0)\approx 2.2$ MHz, the slow decoherence studied above essentially acts as a transfer channel onto the attractive potential surface. For much larger decoherence rates, we observe a total arrest of the acceleration of the dimer through dipole-dipole interactions as shown in \fref{zeno_arrest}. For these large $\gamma$, the coherence between $\ket{\pi_{1}}$ and $\ket{\pi_{2}}$ is quickly damped to zero following switch-on of controllable decoherence, as seen in \frefp{zeno_arrest}{a}. It is subsequently forced to remain zero. For $\sub{\rho(r,r')}{12}=\sub{\rho(r,r')}{21}\equiv0$ one can see from \eref{motionalmasterequation} that no acceleration will take place, since all terms $\sim W_{12}(r)$ vanish. This reflects that dipole-dipole interactions essentially rely on the coherence between the two basis states $\ket{\pi_{1}}$ and $\ket{\pi_{2}}$. To quantify the arrest of acceleration, we show in \frefp{zeno_arrest}{b} the normalized final kinetic energy  $E_f=E(t_f)/E_0$~\cite{footnote:zeropoint} of the dimer. At the chosen final time, $t_f=19.8{\mu}s$, the dimer has been significantly accelerated to kinetic energy $E_0$ in the fully coherent case with $\gamma=0$.
For rates $\gamma$ in excess of $W_{12}(r_0)\approx 2.2$ MHz, which are in reach of experiments \cite{sup:info}, acceleration is almost entirely suppressed.

We recognize a quantum Zeno effect, in which a system that is sufficiently frequently measured is frozen in its initial quantum state \cite{misra:quantumzeno,Kofman:antizeno}. Since the dephasing rate $\gamma$ arises through gathering position and state information about our embedded Rydberg dimer \cite{schoenleber:immag}, the phenomenon shown in \fref{zeno_arrest} furnishes a position space manifestation of the quantum Zeno effect, intriguingly close in spirit to the original philosophy of Zeno regarding the motion of an arrow.

\ssection{Control through decoherence}
%
Employing decoherence and dissipation to engineer quantum states is becoming an active field of research \cite{verstraete:dissipquantcomp,Tiersch:Decoherence:magnetoreception,schempp:spintransport,cenap:dissipbind,Lemeshko:dissipbind}. The preceding example already demonstrates a type of control over the motional state through decoherence. An attractive feature of controllable decoherence due to EIT imaging in our setup is the possibility to vary it in time. We now consider a time-dependent dephasing rate $\gamma(t) = \sum_n^P \gamma_0 \exp{[-(t-t_{n0})^2/\tau^2]}$, which is the sum of $P$ Gaussian pulses at times $t_{n0}$ with durations $\tau$.

\begin{figure}[htb]
\centering\epsfig{file=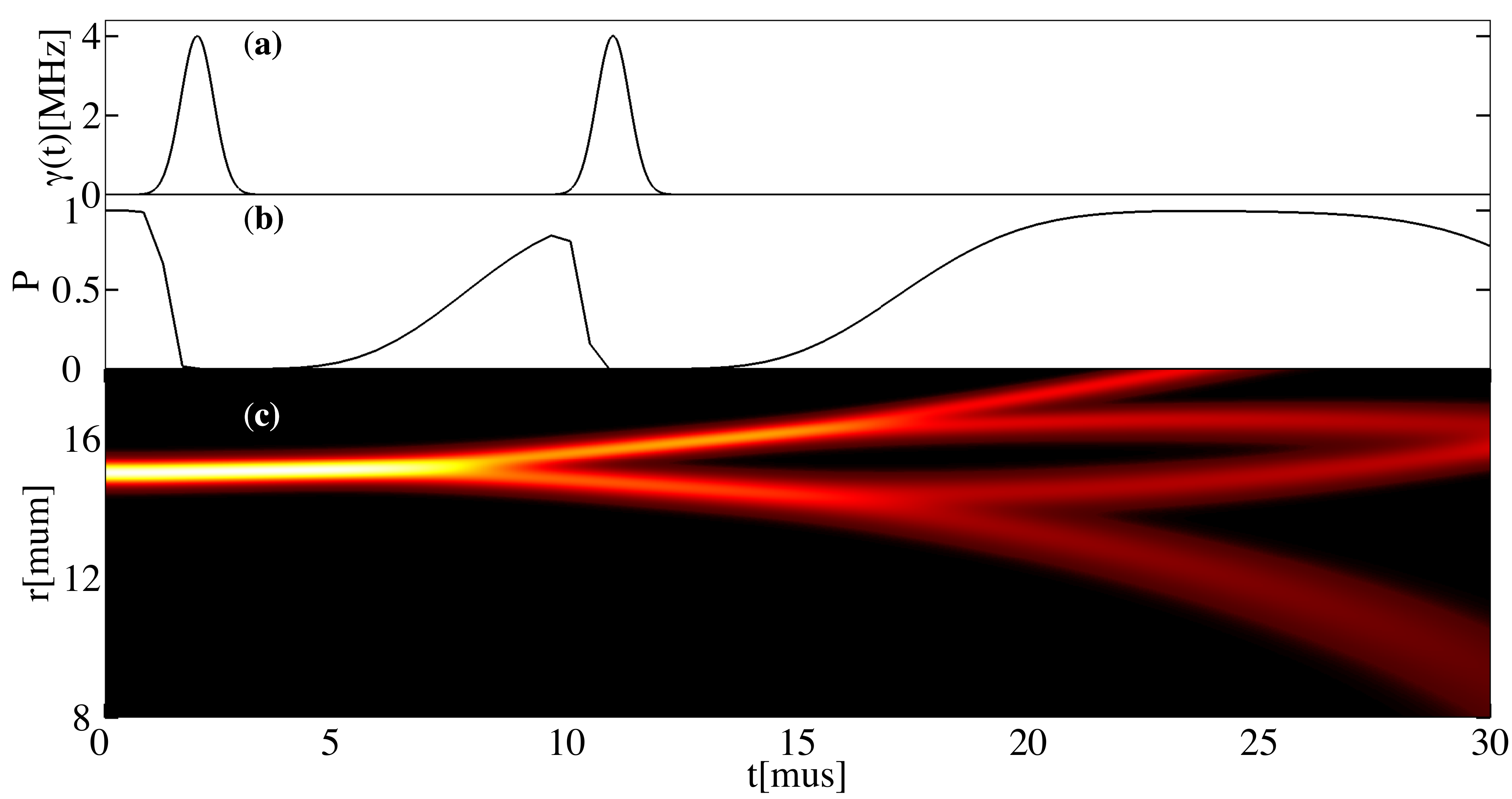,width= \figurewidth\columnwidth} 
\caption{(color online) Repeated application of dephasing pulses with $\gamma_0=4$ MHz can control relative system populations on the repulsive- and attractive energy surfaces. (a) Time dependent dephasing rate $\gamma(t)$. (b) Resulting mean local purity $\sub{\bar{P}}{loc}(t)$, see text. (c) Ensuing radial probability density $n(r,t)$.
\label{decoherence_pulses}}
\end{figure}
We show in \fref{decoherence_pulses} that strong pulses with $\gamma_0=4$ MHz can suddenly redistribute population from one energy surface onto both in an incoherent fashion. The simulation begins in the pure repulsive initial state \bref{rhoini}. The decohering pulse around $t_{10}=2$ $\mu$s rapidly turns this into the mixed state $\hat{\rho}_M$, which contains both surfaces.  Due to dipole-dipole forces, the populations of the dimer on the repulsive and attractive surfaces spatially de-mix with some delay after the pulse, as can be seen in panel (c) around $10$ $\mu$s. This is also reflected in mean local purity, defined as $\sub{\bar{P}}{loc}(t) = \int dr P(r,t) n(r)$ with $P(r,t) = \{[\sub{\rho(r,r)}{11}^2 +  2 \sub{\rho(r,r)}{12}\sub{\rho(r,r)}{21} +  \sub{\rho(r,r)}{22}^2]/[\sub{\rho(r,r)}{11}^2  +  \sub{\rho(r,r)}{22}^2]\}-1$. We have chosen $P(r,t)$ such that $P=1$ when the electronic state at distance $r$ is pure and $P=0$ when it is mixed. The spatial demixing thus leads to a revival of the mean local purity. The dimer-components on the repulsive and attractive surface are then de-cohered a second time around $t_{20}=11$ $\mu$s, resulting, again with some delay, in a total of four separate component of motion. Once these have spatially segregated ($t\approx23$ $\mu$s ), the mean local purity returns to unity, since locally the system is again everywhere in a pure state. 

\ssection{Practical implementation}
 %
The Rydberg electron experiences direct contact interactions with surrounding ground-state atoms, in addition to the controllable interaction via EIT \cite{greene:ultralongrangemol}. This can cause a reduction of Rydberg state life-times \cite{niederpruem:giantion,balewski:elecBEC} and residual de-coherence. Since a rate $\gamma=1$ MHz can be realized at relatively low densities $\rho=5\times10^{17}$ m$^{-3}$ \cite{sup:info}, with a mean number of only $0.25$ ground-state atoms in the Rydberg orbital volume, we expect these effects to be small, with an extrapolated Rydberg \emph{dimer} life-time of about $27$ $\mu$s. Also (dressed) forces acting on ground-state atom are small. We can calculate the strength of a disorder term $\Delta E$ in \eref{motionalmasterequation} as discussed in \cite{schoenleber:immag} and reach $\Delta E\sim0.1$ MHz for the parameters above, small compared to dipole-dipole strengths $W_{12}$.  

\ssection{Conclusions and outlook}
%
We have shown how forces controlling a dimer of Rydberg atoms can be gradually turned from repulsive to attractive or entirely suppressed, through controlled destruction of the superposition of two-atom electronic states that furnish the underlying molecular potential. Our results highlight the importance of intra-molecular quantum coherence for the definition of chemical Born-Oppenheimer surfaces, and open up a research arena on the influence of controllable decoherence \cite{schoenleber:immag} on energy transport and atomic motional dynamics in flexible Rydberg aggregates \cite{cenap:motion,moebius:cradle,wuester:cradle,wuester:CI,leonhardt:switch} embedded in host atom clouds \cite{moebius:bobbels,wuester:cannon}. The most extreme form of decoherence in dipole-dipole interactions leads to a quantum-Zeno effect in the spatial acceleration of the Rydberg dimer. Finally we have shown how temporally localized decoherence pulses can be used as an incoherent means of population re-distribution among BO surfaces.

In the present work we consider parameters where the disorder potential ${\Delta E}$ of the dimer due to its interaction with randomly located background atoms can be neglected. For other choices of parameters the disorder potential can be made dominant over dephasing \cite{schoenleber:immag}, suggesting an accessible model system for quantum motion in disordered potentials. The combination of decoherence channels discussed here with intersecting BO surfaces \cite{wuester:CI,leonhardt:switch} may allow accessible laboratory model studies of quantum chemical phenomena such as relaxation across a conical intersection \cite{Perun-Domcke-Abinitiostudies-2005}.

\acknowledgments

We gladly acknowledge interesting discussions with Alexander Eisfeld, Michael Genkin and Shannon Whitlock, and  EU financial support received from the Marie Curie Initial Training Network (ITN) COHERENCE".


\section{Supplemental information}

This supplemental material provides the derivation of our master equation governing dimer motion and the complete Hamiltonian describing the dimer embedded in a background cold gas.

\ssection{Master equation for relative motion}
We intend to treat the relative coordinate $r$ of the dimer quantum mechanically. Due to the presence of decoherence in the system, we have to describe it with a density matrix 
\begin{align}
\hat{\rho}=\sum_{\stackrel{kl}{ab}} \rho_{ka,lb} \ket{\phi_k}\otimes \ket{\pi_a}  \bra{\phi_l}\otimes \bra{\pi_b},
\label{densitymatrix}
\end{align}
where $\ket{\pi_i}$ is the basis of the electronic state and $\ket{\phi_n}$ an arbitrary basis for the motional Hilbertspace. From the Hamiltonian (1) in the main article, we derive the van-Neumann equation $i\hbar\dot{\hat{\rho}}=[\hat{H},\hat{\rho}]$ with the short-hand $\hat{T}=-\hbar^2 \partial^2/\partial r^2$ for the kinetic energy operator. The resulting expression is immediately sandwiched between $\bra{r}\otimes\bra{\pi_n} \dots \ket{r'}\otimes\ket{\pi_m} $, where $\ket{r}$ is the abstract position space basis, so that $\phi(r)_k = \braket{r}{\phi_k}$ is the position space representation of the basis element $\ket{\phi_k}$. We obtain:
\begin{align}
&i\hbar \sum_{kl}\dot{\rho}_{kn,lm}\braket{r}{\phi_k}\braket{\phi_l}{r'}=
\CR
&= \sum_{kl}\bigg\{
\rho_{kn,lm} \left(\bra{r}\hat{T}\ket{\phi_k}\braket{\phi_l}{r'} -\braket{r}{\phi_k}  \bra{\phi_l}\hat{T}\ket{r'}\right)
\CR
&\sum_{i}W_{12} \left( \rho_{ki,lm}  \bar{\delta}_{in} -  \rho_{kn,li} \bar{\delta}_{im} \right)\braket{r}{\phi_k}\braket{\phi_l}{r'}
\bigg\},
\label{step2}
\end{align}
where $ \bar{\delta}_{ij}=1-\delta_{ij}$, using $i,j\in\{1,2\}$, and $\delta_{ij}$ is the Kronecker delta.
We now define the position space representation of the density matrix $ \sub{\rho(r,r')}{nm}= \sum_{kl} \rho_{kn,lm}\braket{r}{\phi_k}\braket{\phi_l}{r'}$ used in the main article and note that
 $-\hbar^2 \partial^2/\partial r^2 \sub{\rho(r,r')}{nm}= \sum_{kl} \rho_{kn,lm}\bra{r}\hat{T} \ket{\phi_k}\braket{\phi_l}{r'}$ and  $-\hbar^2 \partial^2/\partial r'^2 \sub{\rho(r,r')}{nm}= \sum_{kl} \rho_{kn,lm}\braket{r}{\phi_k}\bra{\phi_l}\hat{T}\ket{r'}$
are the consistent position space representations of the kinetic energy operator. We finally arrive at the first two lines of 
Eq.~(2) of the main article. The third line in the main article arises through coupling to the background gas and is discussed in the next section.

\ssection{Hamiltonian for embedded dimer system}
We describe the detailed interplay of Rydberg atoms embedded in (and interacting with) an optically driven background gas in \cite{schoenleber:immag}. Here we provide a brief summary of the relevant Hamiltonian and results. The entire system of Rydberg dimer \emph{and} background atoms is governed by a many-body Masterequation
\begin{align}
\sub{\dot{\hat{\rho}}}{MB}= -i [\hat{H},\sub{\hat{\rho}}{MB}] + \sum_\alpha {\cal L}_{\hat{L}_\alpha}[\sub{\hat{\rho}}{MB}].
\label{mastereqn}
\end{align}
Here $\sub{\hat{\rho}}{MB}$ now describes only the electronic state space of dimer \emph{and} background atoms, the motion of the dimer has been separately discussed above. 
This disentangling of the motional- and part of the electronic dynamics is justified when the equilibration time-scale inherent in \bref{mastereqn} is much faster than that of motion, which is fulfilled here.

The Hamiltonian consists of four parts, $\hat{H}=\sub{\hat{H}}{dd}+\sub{\hat{H}}{EIT}+\sub{\hat{H}}{int}$, for the dipole-dipole interactions, the 
background gas of three-level atoms and van-der-Waals (vdW) interactions \cite{book:gallagher,singer:VdWcoefficients} between atoms that are in a Rydberg state. The super-operator ${\cal 
L}_{\hat{L}_\alpha}[\hat{\rho}]$ describes spontaneous 
decay of the background atom $\alpha$ from level $\ket{e}$, thus 
 $ {\cal L}_{\hat{O}}[\hat{\rho}]=\hat{O}\hat{\rho}\hat{O}^\dagger - (\hat{O}^\dagger\hat{O}\hat{\rho}+\hat{\rho}\hat{O}^\dagger\hat{O})/2$ and the decay 
operator is $\hat{L}_\alpha=\sqrt{\Gamma_p}\hat{\sigma}^{(\alpha)}_{ge}$, with $\hat{\sigma}^{(\alpha)}_{kk'}=[\ket{k}\bra{k'}]_\alpha$ acting on atom $\alpha$ 
only and $k,k'\in \{g,e,r,s,p\}$.
  
In addition to $\sub{\hat{H}}{dd}$ from the main article the Hamiltonian for the background gas in the rotating wave approximation reads 
\begin{align}
\sub{\hat{H}}{EIT}= \sum_\alpha  &\bigg[\frac{\sub{\Omega}{p}}{2}\hat{\sigma}^{(\alpha)}_{eg}  + \frac{\sub{\Omega}{c}}{2}\hat{\sigma}^{(\alpha)}_{re}  + \mbox{h.c.}  \bigg],
\label{Heit}
\end{align}
where $\Omega_{p,c}$ are the probe and coupling Rabi frequencies. Typically $\Omega_p \ll \Omega_c$ which corresponds to conditions of electromagnetically induced transparency (EIT) used for Rydberg atom detection \cite{guenter:EIT,guenter:EITexpt}.

Background atoms interact among themselves and with the aggregate through vdW interactions
\begin{align}
\sub{\hat{H}}{int}&= \sum_{\alpha<\beta} V_{\alpha\beta}^{(rr)} \hat{\sigma}^{(\alpha)}_{rr}\hat{\sigma}^{(\beta)}_{rr} +  \sum_{a\in\{s,p\},\alpha n } \!\! V_{\alpha n}^{(ra)} \hat{\sigma}^{(\alpha)}_{rr}\hat{\sigma}^{(n)}_{aa}.
\label{Hint}
\end{align}
As in \cite{schoenleber:immag} we assume isotropic interactions, and quantum states $\ket{s}=\ket{43s}$, 
$\ket{p}=\ket{43p}$ and $\ket{r}=\ket{38s}$ in $^{87}$Rb, where $V_{\alpha\beta}^{(rr)}=C_{6,rr}/|\bv{x}_\alpha-\bv{x}_\beta|^6$, $V_{\alpha n}^{(rs)}=C_{6,rs}/|\bv{x}_\alpha-\bv{x}_n|^6$ and $V_{\alpha n}^{(rp)}=C_{4,rp}/|\bv{x}_\alpha-\bv{x}_n|^4$ \cite{footnote:interactions}.

As discussed in \cite{schoenleber:immag}, we can adiabatically eliminate the dynamics of all background atoms if the time-scale on which their electronic dynamics approaches a steady state ($1/\Gamma_p$) is faster than the time-scale of electronic dynamics in the embedded dimer $1/W_{12}$, which will be fulfilled here.

We then arrive at an effective equation of motion for the dimer only, which for the dimer discussed here is
\begin{align}
\dot{\rho}_{nm}= \sum_{k} i (W_{km} \rho_{nk} - W_{nk} \rho_{km}) 
+ \left( i \Delta E- \frac{\gamma}{2}\right) \rho_{nm}, 
\label{effME}
\end{align}
where $\Delta E = E_1 - E_2 + \epsilon_{12}$
with
$E_n = \sum_\alpha \sub{E}{eff}^{(n\alpha)}$, $\epsilon_{12}= \sum_\alpha\mbox{\cal Im}[\sub{L}{eff}^{(1\alpha)}\sub{L}{eff}^{(2\alpha)*}]$, and
$ \gamma =  \sum_\alpha(|\sub{L}{eff}^{(1\alpha)}|^2 + |\sub{L}{eff}^{(2\alpha)}|^2 -2\mbox{\cal Re}[\sub{L}{eff}^{(1\alpha)}\sub{L}{eff}^{(2\alpha)*}] )$, 
and
\begin{align}
\sub{E}{eff}&=  \sum_{n}\left[ \sum_\alpha \frac{ \Omega_p^2}{ \Omega_c^2 } \frac{    \bar{V}_{n\alpha} }{1 +    (\bar{V}_{n\alpha}/\sub{V}{c})^2   }   \right],
\label{effective_hamil}
\\
\sub{L}{eff}^{(\alpha)}&= \sum_n \left[  \frac{ \Omega_p}{ \sqrt{\Gamma_p} } \frac{1 }{i +    \sub{V}{c}/\bar{V}_{n\alpha}  } \right].
\label{effective_dephasing}
\end{align}
In these expressions indices $\alpha$ enumerate the background gas atoms, and $n\in\{1,2\}$ the dimer atoms. We used the shorthand $\bar{V}_{n\alpha} = V^{(rp)}_{n\alpha}+ \sum_{m\neq n} V^{(rs)}_{m\alpha}$. 
The quantity $\Delta E$ controls the energy mismatch between the two electronic states of the dimer ($\ket{\pi_{1,2}}$), due to the slightly different interaction of the dimer atoms with surrounding Rydberg-EIT-dressed background atoms. 
Meanwhile $\gamma$ is the corresponding measurement induced decoherence, since a monitoring of the optical response of background now allows us to infer if the dimer state is $\ket{\pi_1}$ or  $\ket{\pi_2}$. For a more in-depth discussion  of these effects see \cite{schoenleber:immag}.

\begin{figure}[htb]
\centering\epsfig{file=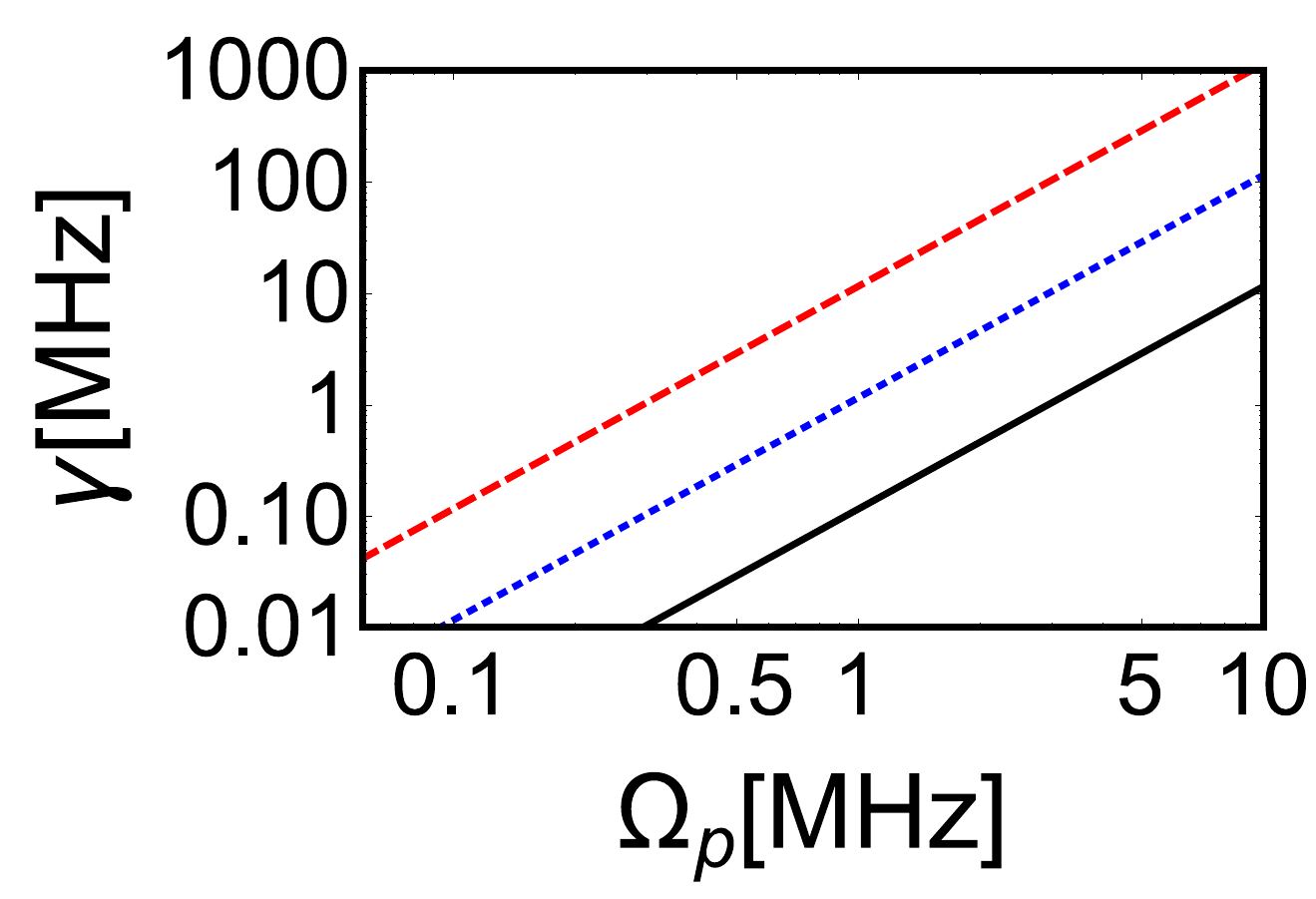,width= 0.7\columnwidth} 
\caption{(color online) Achievable dephasing rates, for densities $\rho=10^{16}m^{-3}$ (black),  $\rho=10^{17}m^{-3}$ (blue dotted),  $\rho=10^{18}m^{-3}$ (red dashed). Other parameters were $\Omega_c=30$ MHz, $\Gamma_p=6.1$ MHz. We see that all values refereed to in the main article can be engineered. 
\label{gamma_ranges}}
\end{figure}

In the limit of a dense distribution of background atoms, we have analytically calculated the dephasing rate $\gamma$ in the supplemental information of \cite{schoenleber:immag}.
We show achievable values of $\gamma$ for various realistic background gas densities and $\Omega_p$ in \fref{gamma_ranges}.
%

\ssection{Kinetic and potential energies}
For the data presented in Fig.~3 of the main article we require the kinetic energy $\sub{E}{kin}$ of the Rydberg dimer, described by the density matrix $\hat{\rho}$. Using the same techniques discussed in the first section of this supplemental material, we arrive at:
\begin{align}
\sub{E}{kin}= \mbox{Tr}\left[\frac{\hat{p}^2}{2m}\hat{\rho}\right] = \sum_n \int dk \left(\frac{\hbar^2 k^2}{2m} \right)\tilde{\rho}(k,-k)_{nn}.
\label{Ekin} 
\end{align}
We have expressed this in terms of the Fourier-transform $\tilde{\rho}(k,k')_{nm} =\int dr \int dr' \exp{[-i(kr+k'r')]} \rho(r,r')_{nm}/(2\pi)$ of the dynamical variables in Eq.~(2) of the main text. Similarly the potential energy is
\begin{align}
\sub{E}{pot}= \mbox{Tr}\left[\hat{H}_{dd}(\hat{r})\hat{\rho}\right] =  \int dr W_{12}(r)\left(  \rho(r,r)_{12} +  \rho(r,r)_{21} \right) .
\label{Epot} 
\end{align}
%


\begin{thebibliography}{65}
\bibitem[{foo()}]{footnote:zeropoint}
\bibinfo{note}{We take the final kinetic energy
  $\sub{E}{kin}=$Tr$[\hat{\rho}(-\hbar^2 \nabla^2/m)]$ at $t=t_f$ and substract
  the initial zero point energy ($\sub{E}{kin}$ at $t=0$).}
\bibitem[{foo()}]{footnote:interactions}
\bibinfo{note}{For these quantum numbers we obtain dispersion coefficients
  $C_{6,rr}=3.67 \times 10^{18}$ au, $C_{6,rs}=-6 \times 10^{17}$ $C_{4,rp}=-2
  \times 10^{10}$ from exact diagonalisation of the di-atomic Hamiltonian.
  Where comparable these agree with \cite{singer:VdWcoefficients}. The large
  difference of the latter two arises due to a nearly resonant process: $\nu
  p+\nu' s \leftrightarrow (\nu-2)d+\nu'p$. Due to this resonance $V_{rp}$
  contains several dispersion coefficients, the $R^{-4}$ form has been chosen
  to reproduce the same critical radius $R_{c,p}$ as the full interaction for
  our parameters.}
\end{thebibliography}
\end{document}